\documentclass[twocolumn]{aastex63}

\hypersetup{linkcolor=red,citecolor=blue,filecolor=cyan,urlcolor=magenta}

\usepackage[]{units}
\usepackage{gensymb}
\usepackage{amsmath}
\usepackage{booktabs}
\usepackage{romannum}
\usepackage{longtable}
\usepackage{enumitem}

\newcommand{\hi}{\ion{H}{1}}
\newcommand{\hii}{\ion{H}{2}}

\newcommand{\nii}{\ion{N}{2}}

\newcommand{\ha}{H\ensuremath{\alpha}}

\newcommand{\kms}{\mathrm{km~s}\ensuremath{^{-1}}}
\newcommand{\vlsr}{\ensuremath{v_\mathrm{LSR}}}

\makeatletter
\newcommand*{\rom}[1]{\expandafter\@slowromancap\romannumeral #1@}
\makeatother

\received{2020 May 29}

\submitjournal{ApJL}

\shorttitle{\ha\ Towards Fermi Bubble}
\shortauthors{Krishnarao et al.}

\graphicspath{{./}{figures_eps/}} 

\begin{document}

\title{Discovery of High-Velocity H$\alpha$ Emission in the Direction of the Fermi Bubble} 

\correspondingauthor{Dhanesh Krishnarao}
\email{krishnarao@astro.wisc.edu}

\author[0000-0002-7955-7359]{Dhanesh Krishnarao}
\affiliation{Department of Astronomy, University of Wisconsin-Madison, Madison, WI, USA}

\author[0000-0002-8109-2642]{Robert A. Benjamin}
\affiliation{Department of Physics, University of Wisconsin-Whitewater, Whitewater, WI, USA}

\author[0000-0002-9947-6396]{L. Matthew Haffner}
\affiliation{Department of Physical Sciences, Embry-Riddle Aeronautical University, Daytona Beach, FL, USA}

\begin{abstract}

Wisconsin H-Alpha Mapper (WHAM) observations reveal high-velocity \ha\ and [\nii]$\lambda6584$ emission lines in the same direction and velocity as ultraviolet absorption-line features that have been previously associated with the biconical gamma-ray lobes known as the Fermi Bubbles. 
We measure an extinction-corrected intensity of $I_{\ha}=\unit[0.84^{+0.10}_{-0.09}]{Rayleigh}$ for emission with line center $\vlsr=\unit[-221\pm3]{\kms}$, corresponding to an emission measure of $EM = \unit[2.00^{+0.64}_{-0.63}]{\textrm{cm}^{-6}~\textrm{pc}}$.
This emission arises at the same velocity as Hubble Space Telescope/Cosmic Origins Spectrograph observations of ultraviolet absorption features detected in the PDS~456 quasar sight line that passes through the northern Bubble near $l = 10\fdg4, b = +11\fdg2$.
We estimate the total column density of ionized gas in this velocity component to be $N(H^{+}) = \unit[\left(3.28 \pm 0.33\right) \times 10^{18}]{\textrm{cm}^{-2}}$.
The comparison of ionized gas emission and absorption yields an estimate for the characteristic density of $n_{e,c} = \unit[1.8 \pm 0.6]{\textrm{cm}^{-3}}$ and a characteristic length of $L_{c} =\unit[0.56 \pm 0.21]{pc}$ assuming $30\%$ solar metallicity.
For a temperature of $T_{e}=\unit[8500^{+2700}_{-2600}]{K}$---consistent with the measured line widths and [\nii]/\ha\ line ratio---the gas has a thermal pressure of $\nicefrac{p}{k} = \unit[32,000^{+15,000}_{-14,000}]{\textrm{cm}^{-3}~K}$.
Assuming the gas is $\sim $ \unit[6.5]{kpc} distant, the derived density and pressure appear to be anomalously high for gas $\sim \unit[1.3]{kpc}$ above the Galactic midplane. The large thermal pressure is comparable to both a hot halo or Fermi Bubble model, but suggest that the \ha\ arises in an overpressurized zone.

\end{abstract}

\section{Introduction} \label{sec:intro}

The past two decades have seen a resurging interest in the existence of an outflow from the nucleus of the Galaxy. Observations of soft X-rays and radio continuum \citep{Snowden1997, Sofue2000, Almy2000} and later in the mid-infrared and hard X-rays \citep{JBH2003} were interpreted as emission from gas in the nucleus and bulge of the Galaxy, as were microwave observations of the ``Wilkinson Microwave Anisotropy Probe (WMAP) haze" \citep{Finkbeiner2004, Dobler2008}. The discovery of the ``Fermi Bubbles", large lobes extending $\sim \unit[55]{\degree}$ above and below the Galactic midplane visible in gamma-ray emission \citep{Su2010, Dobler2010, Ackermann2014}, has fueled even more interest in the possibility of Galactic nuclear outflows. Other continuum observations, e.g. polarized synchrotron radiation at radio wavelengths \citep{Carretti2013}, have been interpreted in the context of these gamma-ray results. These observations only provide a $2$D image, but their location on the sky toward Galactic Center and relative symmetry across the Galactic plane are argued to support the hypothesis that the emission arises from the center of the Galaxy. Additionally, biconical outflows driven from galactic nuclei are also found in extragalactic systems \citep{Bland1988, Cecil2001, Veilleux2002}, where supernovae, stellar winds, or active galactic nuclei can also power large-scale galactic winds \citep[see][for review]{Heckman2002,Veilleux2020}. 

Of particular interest are spectroscopic observations to measure the kinematics of gas associated with the $2$D footprint of the Fermi Bubbles. $21$-cm \hi\ observations have shown evidence for neutral gas consistent with a constant velocity Galactic Center outflow \citep{McClure-Griffiths2013, DiTeodoro2018, Lockman2020} within $\unit[10]{\degree}$ of the midplane. Quasar absorption-line spectroscopy was also used to detect signatures of outflowing and infalling gas at high latitudes above and below Galactic Center \citep{Keeney2006}. Later, \citet{Fox2015} used the Hubble Space Telescope (HST) to provide evidence for UV absorption closer to the plane that was interpreted as arising from the front and back side of the northern Fermi Bubble along a single line of sight toward the quasar PDS~456 ($l = 10\degree.4, b = +11\degree2$). This sight line passes through the $\unit[1.5]{keV}$ emission footprint from \citet{Snowden1997} and does not contain any known high-velocity \hi\ emission \citep[][constrained using the Green Bank Telescope]{Fox2015}. Since then, more pencil-beam UV sight lines have been studied toward both the northern and southern Fermi Bubbles \citep[e.g.][]{Bordoloi2017, Karim2018}. 

\citet{Miller2016} used XMM-Newton and Suzaku X-ray observations of \ion{O}{7} and \ion{O}{8} emission to model and constrain the properties of a hot gas halo interacting with a large-scale outflow. They found that the energetics involved were consistent with the bubbles forming from a nuclear accretion event at Sgr A*, as opposed to a central starburst. More recently, \citet{JBH2019} used HST UV absorption-line ratios from the Magellanic stream to infer the presence of a Seyfert explosion $\sim \unit[3.5]{Myr}$ ago, emitting large amounts of ionizing radiation into the conical region of the present-day Fermi Bubbles.

Here, we present new Wisconsin H-Alpha Mapper (WHAM) observations of \ha\ and [\nii] towards the quasar PDS~456, where high-velocity UV absorption-lines have previously been seen with HST/Cosmic Origins Spectrograph \citep[COS][]{Fox2015}. These optical spectra provide a new avenue to constrain both the physical conditions of the ionized gas that has been associated with the Fermi Bubbles as well the radiation field emerging from the Galactic Center region and within the Fermi Bubbles. In Section~\ref{sec:data}, we briefly describe the optical and UV observations. Section~\ref{sec:methods} explains our analysis methods and  presents our new \ha\ and [\nii] spectra and the inferred kinematic and physical properties. We discuss these results in Section~\ref{sec:disc} and summarize our conclusions in Section~\ref{sec:conc}. 
All data and python notebooks to replicate the results and figures shown are available in the GitHub repository Deech08/WHAM\_PDS456\footnote{\href{https://github.com/Deech08/WHAM_PDS456}{\texttt{Deech08/WHAM\_PDS456}}}.

\section{Data} \label{sec:data}

\subsection{Optical Spectra}
\begin{figure}[h]
\epsscale{1.2}
\plotone{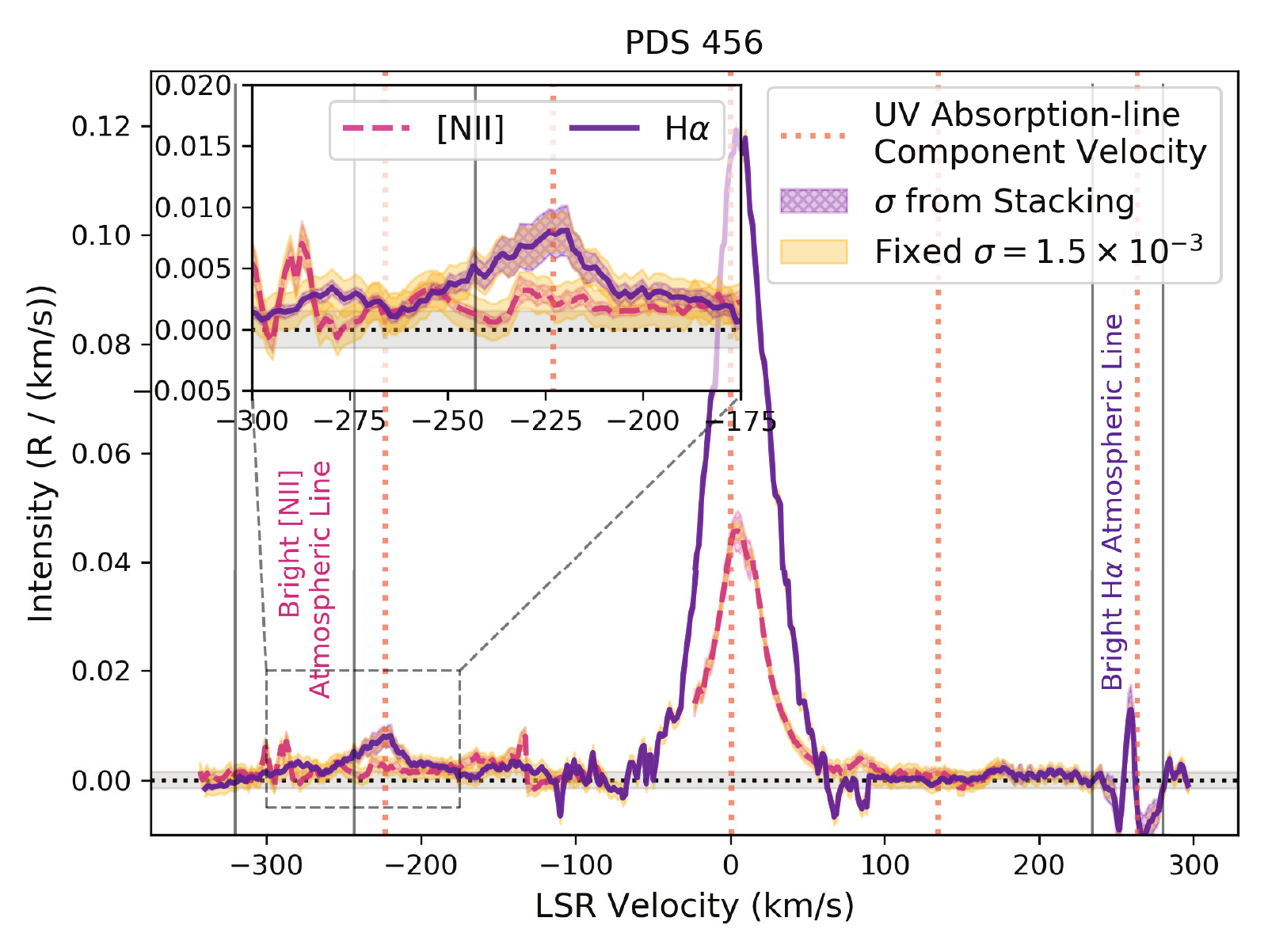}
\caption{Mean stacked \ha\ (purple) and [\nii] (red) emission spectra toward PDS~456 from bootstrap resampling. The yellow shading around the spectra encompasses a range of $\pm \unit[0.0015]{R/(\kms)}$ while the purple/red hatched shading encompasses $1\sigma$ errors from bootstrap resampling. A dotted black line and shading run across the $\unit[0 \pm 0.0015]{R/(\kms)}$ baseline. Dotted orange vertical lines are at the median velocities where UV absorption-line spectra are found at $\vlsr = \unit[-223, 0.5, 134.5, \text{and }263]{\kms}$ \citep{Fox2015, Bordoloi2017}. High-velocity \ha\ and [\nii] emission is detected near the $\vlsr = \unit[-223]{\kms}$ absorption component. The bright emission near $\vlsr = \unit[0]{\kms}$ is from local gas emission. Only \ha\ observations currently extend beyond $\vlsr = \unit[+250]{\kms}$, but a bright atmospheric line strongly contaminates this velocity range of interest. A bright atmospheric line also contaminates the [\nii] spectra between $\unit[-320]{\kms} \lesssim \vlsr \lesssim \unit[-243]{\kms}$, making an accurate measurement of the [\nii] emission difficult in its vicinity. \label{fig:OpticalSpectra}}
\end{figure}

Optical spectra were obtained using WHAM \citep[][see the WHAM-SS release documentation for details\footnote{\href{http://www.astro.wisc.edu/wham/}{http://www.astro.wisc.edu/wham/}}]{wham-nss, wham-south}, currently located at Cerro-Tololo Inter-American Observatory (CTIO) in Chile. Each $30$ -- $120$ second observation yields a $\unit[200]{\kms}$ velocity-range spectrum around \ha\ or [\nii] integrated over a $1\degree$ beam with a velocity resolution of $\sim \unit[11]{\kms}$. 
The deep spectrum presented here in Figure \ref{fig:OpticalSpectra} toward PDS~456 is derived from a total of $\unit[112.5]{minutes}$ of integration time for \ha\ and $\unit[66]{minutes}$ for [\nii] observed during the summer and early fall of 2019. These spectra are processed using \texttt{whampy} \citep{whampy} by applying a flat-field and subtracting an atmospheric template and constant baseline that is fit to the spectra using \texttt{lmfit} \citep{lmfit}. Bright \ha\ sources, such as Sh 2-264 ionized by $\lambda$ Ori \citep{Sahan2016}, are used to apply night-to-night atmospheric corrections to the processed spectra. Several individual observations are then combined using bootstrap resampling to create a continuous, higher signal-to-noise spectrum. Details of the data processing can be found in \citet{wham-nss}. For the deep optical spectra presented here, we estimate a rms error of $\unit[0.0015]{R/(\kms)}$, where $\unit[1]{Rayleigh~(R)} = \unit[\nicefrac{10^6}{4\pi}]{photons~s^{-1}~cm^{-2}~sr^{-1}}$ and corresponds to an emission measure of $\textrm{EM} = \unit[2.25]{cm^{-6}~pc}$ for $T_e = \unit[8000]{K}$ (see Section~\ref{sec:methods:extinct}).

We have additionally begun a mapping campaign of the sky surrounding PDS~456 at high negative velocities. These pilot observations currently consist of $120$-second exposures at each pointing spaced at $\sim \unit[1]{\degree}$ intervals and are processed in the same way as described above, but without the final stacking of several spectra. This results in significantly noisier spectra with rms errors of $\sim \unit[0.01]{R/\kms}$ and will be discussed further in Section~\ref{sec:disc}. 

\subsection{UV Spectra}
\begin{figure}[]
\epsscale{1.15}
\plotone{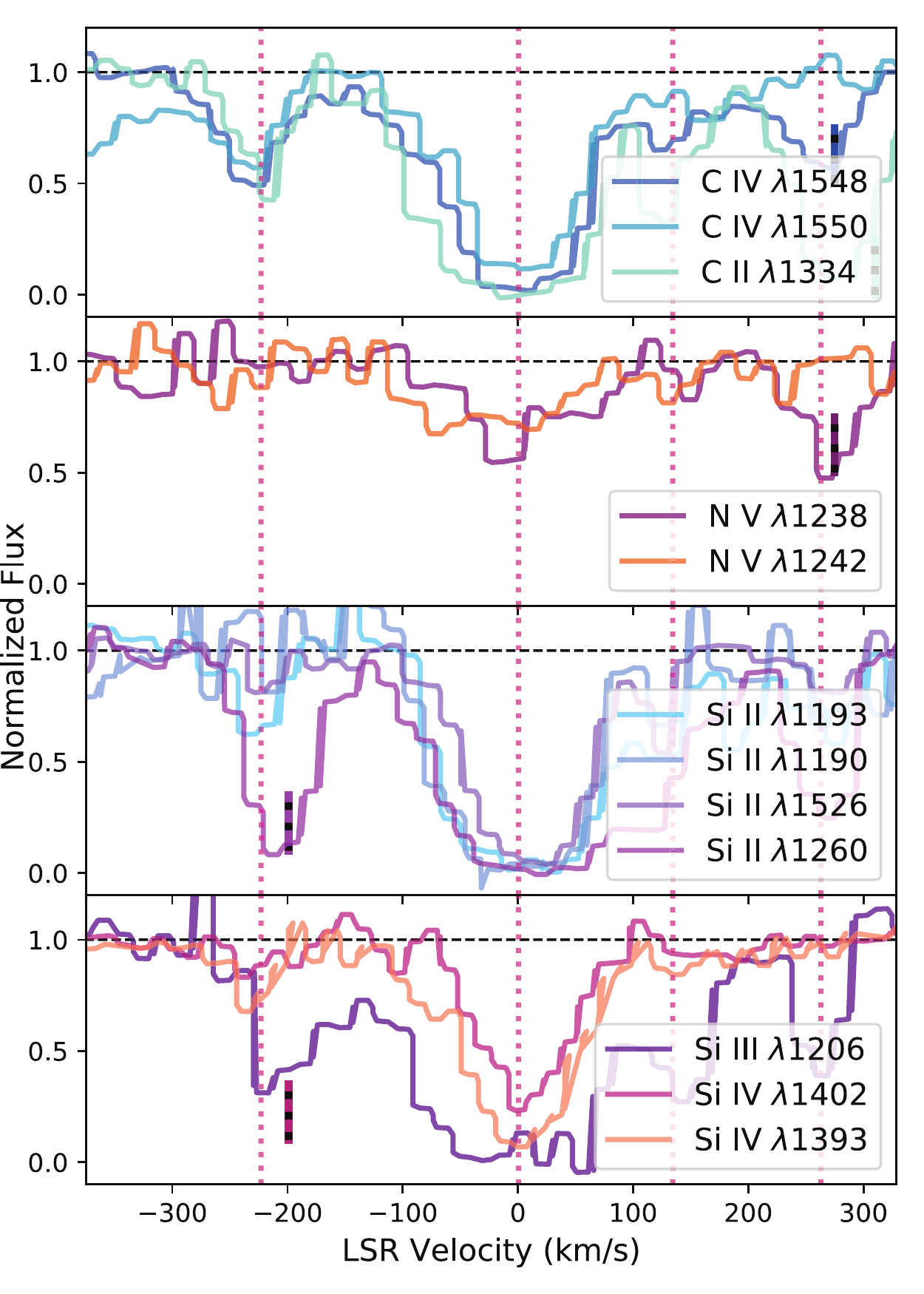}
\caption{Normalized UV absorption-line spectra from HST/COS toward PDS~456, showing different carbon, nitrogen, and silicon ions from top to bottom.. The dotted red vertical lines are at the velocities where UV absorption-line spectra are found as in Figure \ref{fig:OpticalSpectra}. This figure is reproduced from \citet[][their Figure 2]{Fox2015}. Some absorption components, such as the \ion{Si}{3} line near $\unit[-200]{\kms}$ are contaminated by other red-shifted transitions, with these features marked with dashed black vertical lines \citep[see][for details]{Fox2015,Bordoloi2017}\label{fig:UVSpectra}}
\end{figure}

We also use the UV spectra toward PDS~456 originally presented in \citet{Fox2015}, with Voigt-profile fitting results from \citet[][see their Table 2]{Bordoloi2017}. The observations were originally taken in 2014 from COS \citep{Green2012} on board HST using the G130M setting centered on 1291\AA\ and G160M setting centered on 1600\AA. These spectra have an $\mathrm{FWHM} = \unit[20]{\kms}$ spectral resolution with an absolute velocity calibration accurate to within $\unit[5]{\kms}$ and a signal to noise of $\sim \unit[12-20]{\text{per resolution element}}$. For full details on these spectra and their processing, see \citet{Fox2014, Fox2015} and \citet{Bordoloi2017}. In this work, we display UV spectra rebinned to match one resolution element, but the Voigt-profile fitting from \citet{Fox2015} and \citet{Bordoloi2017} is done on the unbinned data. Select spectra are shown in Figure~\ref{fig:UVSpectra}. While there is no detection of \ion{N}{5} absorption, $3\sigma$ upper limits are estimated using the RMS noise of the spectra to be $\log{\left(N_{\textrm{N\rom{5}}~\lambda1238}\right)} < 14.10$ and $\log{\left(N_{\textrm{N\rom{5}}~\lambda1242}\right)} < 14.15$ (Fox 2020, private communication).

\section{Methods and Results} \label{sec:methods}

We use a combination of new optical emission lines and previously measured UV absorption-lines to constrain the temperature, density, and pressure of high-velocity gas features. Throughout this work, uncertainties are reported as $1\sigma$ errors and propagated through calculations using the \texttt{uncertainties} python package\footnote{\href{https://pythonhosted.org/uncertainties/}{https://pythonhosted.org/uncertainties/}} \citep[see also][]{Ku1966}. Here we briefly describe our methods of deriving these constraints.  

\subsection{Optical Line Measurements}\label{sec:methods:optical}
We measure integrated intensities, velocity centroids, and line widths of \ha\ and [\nii] emission lines focusing on the high negative velocity region between $\unit[-270]{\kms} < \vlsr < \unit[-200]{\kms}$. Integrated intensities are computed with the standard ``zeroth moment," while velocity centroids and line widths are derived using the method proposed in \citet{Teague2018}. This method more accurately identifies velocity centroids and uncertainties by fitting a quadratic model to the brightest pixel and its two nearest neighbors in a spectrum. The line width is estimated using a ratio of the peak intensity identified in the above step and the zeroth moment, assuming a Gaussian line profile. This method is preferable to the traditional first and second moments or parameters estimated through Gaussian component fitting because of our relatively low signal to noise. Before finding velocity centroids and line widths, the data is first smoothed using a Savitzky-Golay filter \citep{Savitzky1964} with a width of $\sim \unit[4]{\kms}$. 

The optical spectra are shown in Figure~\ref{fig:OpticalSpectra}, with the velocity region of interest enlarged. The emission near $\vlsr = \unit[0]{\kms}$ is from local emission.
The high negative velocity emission is centered at $\vlsr = \unit[-221 \pm 5]{\kms}$ for \ha\ and $\vlsr = \unit[-230\pm5]{\kms}$ for [\nii], with observed integrated intensities of $I_{\ha} = \unit[0.287 \pm 0.014]{R}$ and $I_{\textrm{N\rom{2}}} = \unit[0.077 \pm 0.011]{R}$ and line widths of $\sigma_v = \unit[14.1\pm2.7]{\kms}$ for \ha\ and $\sigma_v = \unit[9.6\pm4.5]{\kms}$ for [\nii]. 
After correcting for the instrument profile, these line widths are $\sigma_v = \unit[13.3\pm2.7]{\kms}$ for \ha\ and $\sigma_v = \unit[8.4\pm4.5]{\kms}$ for [\nii].
Throughout this work, we use use $\sigma_v$, the standard deviation, to describe line widths.

\subsection{Temperature and Nonthermal Broadening}
We assume the ions observed using both HST/COS (\ion{C}{2}, \ion{Si}{2}, \ion{Si}{4}) and WHAM ([\nii], \ion{H}{2}) are at thermal equilibrium with one another, exhibiting the same gas temperatures and experiencing the same input of turbulence and nonthermal broadening mechanisms. We do not consider the \ion{Si}{3} and \ion{C}{4} absorption-line widths due to contamination from other red shifted transitions or skewed absorption profiles which lead to poor estimates from Voigt profile fits. Then the line widths of these ions are modeled as a function of the ionized gas temperature, $T_e$, mass, $m$, and a nonthermal broadening component, $\sigma_{nonT}$ using
\begin{equation}
\sigma_{model}\left(T_e, m, \sigma_{nonT}\right) = \sqrt{\left(\frac{2 k_B T_e}{m} + \sigma_{nonT}^2 \right)}
\end{equation}

\begin{figure}[h]
\epsscale{1.15}
\plotone{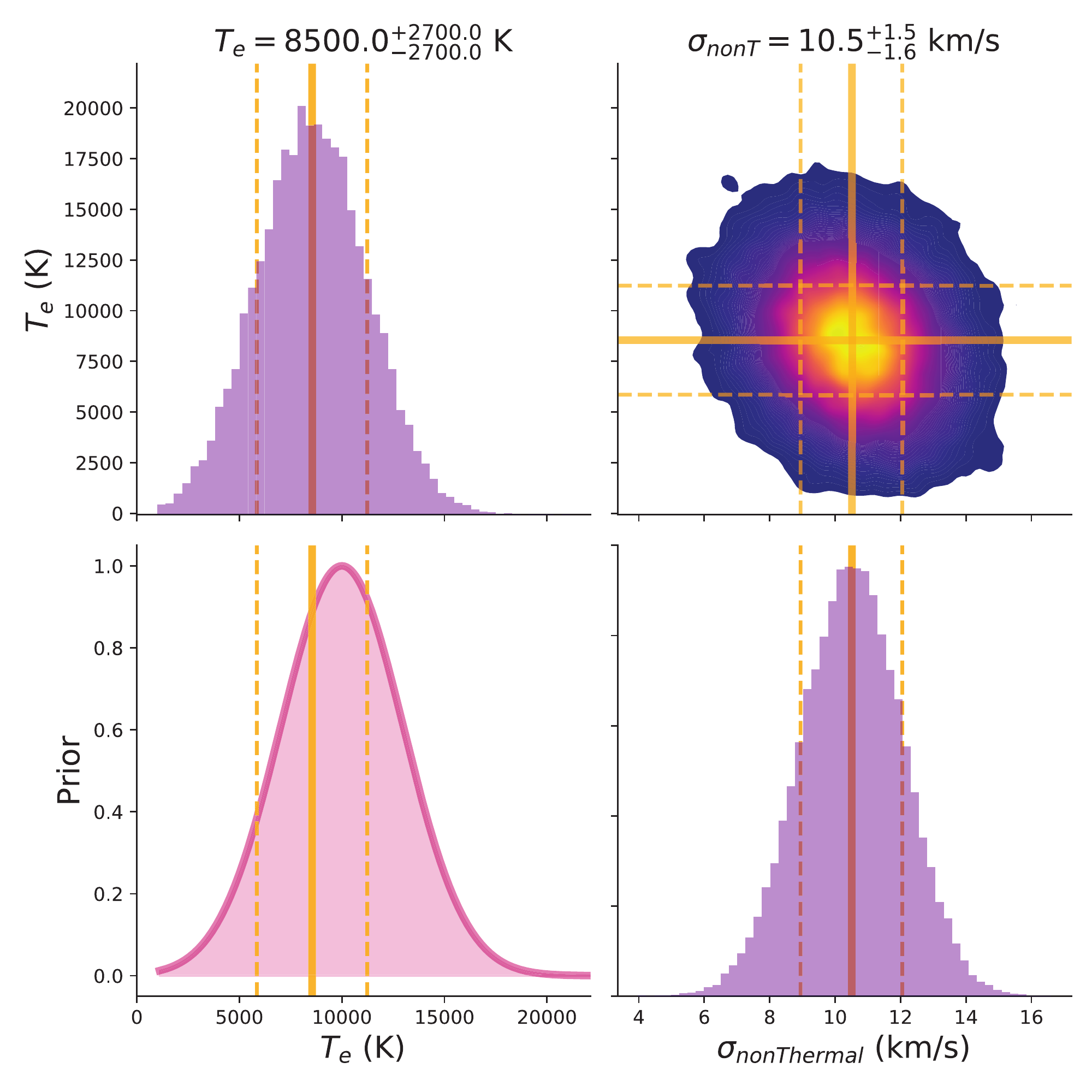}
\caption{Posterior distributions from MCMC fitting of emission line width models assuming a thermal and nonthermal broadening term with a flat prior on nonthermal broadening and a Gaussian prior on the ionized gas temperature. The Gaussian prior for the temperature is shown on the lower left panel. The diagonal panels show marginalized histograms (purple), with the median (solid) and $16$th and $84$th percentile values (dashed) shown in yellow for the gas temperature and nonthermal broadening term. The upper right panel shows a $2$D Gaussian kernel density estimate of both parameters.  \label{fig:MCMC_lineWidths}}
\end{figure}

The warm gas temperature and nonthermal broadening component are constrained using a Bayesian Markov Chain Monte Carlo (MCMC) approach implemented using \texttt{emcee} \citep{emcee}. We use a flat prior on $\sigma_{nonT}$ constrained between $\unit[1]{\kms} < \sigma_{nonT} < \unit[20]{\kms}$ and a Gaussian prior on $T_e$ with a mean of $\unit[10^4]{K}$ and standard deviation of $\unit[3000]{K}$, limited to a range of $\unit[10^3]{K} < T_e < \unit[10^5]{K}$. Our likelihood, $\mathcal{L}$, has the form

\begin{equation} \label{loglike}
\begin{split}
\log{\left(\mathcal{L}\right)} = -\nicefrac{1}{2} \sum_{ions}\left[\frac{\left(\sigma_{obs,ion} - \sigma_{model,ion}\left( m_{ion} \right) \right)^2}{s_{obs,ion}^2}\right. \\
+ \left.\log{\left(s_{obs,ion}^2 \right)}\right]
\end{split}
\end{equation}

where $\sigma_{obs,ion}$ is the observed line width of an ion, $s_{obs,ion}$ is the standard error of the observed line width, and $m_{ion}$ is the atomic mass. Parameter values are derived from the median of the posterior probability distributions after $10,000$ steps, with the $16$th and $84$th percentiles used to estimate errors. The resulting gas temperature and nonthermal broadening contribution are $T_e = \unit[8500^{+2700}_{-2600}]{K}$ and $\sigma_{nonT} = \unit[10.5^{+1.5}_{-1.6}]{\kms}$, with posterior distributions shown in Figure~\ref{fig:MCMC_lineWidths}.

\subsection{Extinction Corrections} \label{sec:methods:extinct}
In order to correct the optical line emission for extinction we use the $3$D dust maps of \citet{Green2019} and the extinction curve of \citet{Fitzpatrick2007}. We adopt the \citet{DiTeodoro2018} kinematic model, which places the emission at a distance of  $D = \unit[6.5 \pm 0.08]{kpc}$. To account for the large uncertainty in constraining actual $3$D distances, we use a distance of $D = \unit[6.5 \pm 0.2]{kpc}$ when estimating $A_V$. We estimate $A_V$ using $10,000$ points distributed uniformly in position within our beam and distributed as a Gaussian around our estimated distance, resulting in $A_V = \unit[1.5 \pm 0.2]{mag}$. The $3$D dust maps predict nearly all of the dust to be located at distances $<\unit[4]{kpc}$, so that the estimated $A_V$ would not change even with distance uncertainties as high as $\pm \sim \unit[2]{kpc}$.

The extinction-corrected intensities are then $I_{\ha} = \unit[0.84^{+0.10}_{-0.09}]{R}$ and $I_{\textrm{N\rom{2}}} = \unit[0.225^{+0.028}_{-0.025}]{R}$. 
The resulting extinction-corrected [\nii]/\ha\ line ratio is $\textrm{[\nii]/\ha} = 0.26 \pm 0.05$. 
Assuming case B recombination and no absorption, \ha\ surface brightness can be related to the emission measure $EM = \int_0^{\infty}{n_e^2 ds}$ as 
\begin{equation}
\begin{split}
    EM = \left(2.77 \textrm{cm}^{-6}~\textrm{pc}\right)~
     \left(\frac{I_{\ha}}{R}\right) \left(\epsilon_{\textrm{b}}\right)^{-1}\\
     \times~T_4^{(0.942 + 0.031~\ln{T_4})}    
\end{split}\label{eq:iha_simp}
\end{equation}
where $T_4 = \nicefrac{T_e}{\unit[10^4]{K}}$, and the constant term is derived from the effective recombination rate of \ha\ \citep{Draine2011} and $\epsilon_{\textrm{b}}$ is a beam dilution factor.
At the inferred temperature of $T_e = \unit[8500^{+2700}_{-2600}]{K}$ and assuming $\epsilon_{\textrm{b}} = 1$, the extinction-corrected emission measure is $EM = \unit[2.00^{+0.64}_{-0.63}]{\textrm{cm}^{-6}~\textrm{pc}}$.
A summary of the measured optical emission and UV absorption-line properties are shown in Table~\ref{tab:measurements}.

\begin{table*}[htb]

\begin{tabular}{lcccccc}
\toprule
Ion                      & Instrument & $\vlsr$    & Intensity     &Intensity        & $\log(N)$          & $\sigma_v$               \\\midrule
                         &            & ($\kms$)       &  $(R)$           &    [Dereddened] $(R)$           & $(\text{cm}^{-2})$ & ($\kms$)                   \\
\hi/$21$-cm$^\textbf{a}$              & GBT       & &      &          &         $<17.48$           & N/A \\                       
\hii/\ha              & WHAM       & $-221 \pm 5$ & $0.287 \pm 0.014$&   $0.84^{+0.10}_{-0.09}$       &                    & $13.3 \pm 2.7 $ \\
$\left[\text{\nii}\right]$  & WHAM       & $-230 \pm 5$ & $0.077 \pm 0.011$& $0.225^{+0.028}_{-0.025}$ &                    & $8.4 \pm 4.5$ \\
\ion{Si}{2}              & HST/COS    & $-223 \pm 2$ &                 &             & $13.02 \pm 0.08$     & $9.3 \pm 2.8$            \\
\ion{Si}{3}$^\textbf{b}$     & HST/COS    & $-197 \pm 2$ &                  &            & $13.13 \pm 0.02$   & $27.8 \pm 1.6$           \\
\ion{Si}{4}              & HST/COS    & $-231 \pm 2$ &                  &            & $12.9 \pm 0.06$    & $13.4 \pm 2.4$           \\
\ion{C}{2}               & HST/COS    & $-220 \pm 6$ &                  &            & $13.8 \pm 0.14$    & $14.1 \pm 5.7$           \\
\ion{C}{4}               & HST/COS    & $-233 \pm 2$ &                  &            & $13.79 \pm 0.03$   & $22.6 \pm 1.5$ \\   
\ion{N}{5}$^\textbf{c}$            & HST/COS   &   N/A     &                       &           & $< 14.10$         &   N/A         \\  \midrule
\hi/$21$-cm$^\textbf{a}$              & GBT       & &      &          &         $<17.48$           & N/A \\                       
\hii/\ha$^\textbf{d}$              & WHAM       & N/A & $<1.24$ &   N/A       &                    & N/A \\
\ion{Si}{2}              & HST/COS    & $264 \pm 2$ &                 &             & $13.37 \pm 0.02$     & $25.1 \pm 1.6$            \\
\ion{Si}{3}              & HST/COS    & $259 \pm 2$ &                  &            & $12.85 \pm 0.04$   & $12.9 \pm 1.6$           \\
\ion{Al}{2}               & HST/COS    & $263 \pm 3$ &                  &            & $13.53 \pm 0.63$    & $0.6 \pm 2.1$           \\   \bottomrule      
\end{tabular}

\begin{flushright}
$^\textbf{a}$ $3~\sigma$ upper limit from \citep{Fox2015} constrained with the Green Bank Telescope (GBT).\\
$^\textbf{b}$ Partial Ly $\beta$ contamination at $z = 0.175539$.\\
$^\textbf{c}$ $3~\sigma$ upper limit\\
$^\textbf{d}$ $3\sigma$ upper limit coincident with a bright atmosphere line (see Section~\ref{sec:meth:upper}).
\end{flushright}
\caption{Measured emission and absorption-line centroids, intensities, column densities, and velocity widths from WHAM (this work) and HST/COS \citep{Fox2015, Bordoloi2017}. Extinction corrections use 3D dust models from \citet{Green2019} and assume a distance of $\unit[6.5 \pm 0.2]{kpc}$.}
\label{tab:measurements}
\end{table*}

\subsection{Ionized Gas Column Density} 

We estimate the ionized gas column density using the silicon absorption-line measurements, assuming that all silicon gas is either singly, doubly, or triply ionized, and at a singe gas temperature. The constrained upper limit on the \hi\ column density toward PDS~456 of $N_{\textrm{H\rom{1}}} < \unit[3.3~\times~10^{17}]{\textrm{cm}^{-2}}$ \citep{Fox2015} supports this fully ionized assumption. The ionized gas column density is then
\begin{equation}
    N_{\textrm{H}^+} = \frac{N_{\textrm{Si\rom{2}}} + N_{\textrm{Si\rom{3}}} + N_{\textrm{Si\rom{4}}}}{\left( \nicefrac{\textrm{Si}}{\textrm{H}} \right)}\label{eq:SiTot}
\end{equation}
where $\left( \nicefrac{\textrm{Si}}{\textrm{H}} \right)$ is the silicon abundance.  

The combination of emission, probing the warm ionized gas density squared, and absorption, probing the ionized gas column density, allows for the ionized gas density to be solved for using 
\begin{equation}
\begin{split}
    n_e = \left( 0.32~\textrm{cm}^{-3}\right)~ \left(\frac{EM}{\textrm{cm}^{-6}~\textrm{pc}}\right)\left(\frac{N_{\textrm{H}^+}}{10^{18}\textrm{cm}^{-2}}\right)^{-1}\label{eq:ne_combo}
\end{split}
\end{equation}
This estimate of $n_e$ has no dependence on the path length, $L$, of emitting/absorbing gas. $L$ can instead be derived as
\begin{equation}
\begin{split}
   L = \left( 9.62~\textrm{pc}\right)~ \left(\frac{N_{\textrm{H}^+}}{10^{18}\textrm{cm}^{-2}}\right)^{2} \left(\frac{EM}{\textrm{cm}^{-6}~\textrm{pc}}\right)^{-1}.\label{eq:L_combo}
\end{split}
\end{equation}
The thermal pressure is approximately $\nicefrac{p}{k} = 2~n_e~T_e$
where the factor of $2$ accounts for the fact that the gas is fully ionized. 

Assuming a solar metallicity of $\left( \nicefrac{\textrm{Si}}{\textrm{H}} \right)_{\odot} = \left( 3.24 \pm 0.22 \right)~\times~10^{-5}$ \citep{Asplund2009} and no depletion onto dust grains, Equation~\ref{eq:SiTot} yields $N_{\textrm{H}^+} = \unit[\left(9.85 \pm 0.99\right)~\times 10^{17}]{\textrm{cm}^{-2}}$. 
Combining with the the extinction-corrected emission measure results in estimates of the characteristic ionized gas density and length of $n_{e} = \unit[6.3 \pm 2.1]{\textrm{cm}^{-3}}$ and $L = \unit[0.05 \pm 0.02]{pc}$.
These in turn provide an estimate of the thermal gas pressure, $\nicefrac{p}{k} = \unit[106,000^{+49,000}_{-48,000}]{\textrm{cm}^{-3}~K}$.

A principal caveat for these estimates is that they are derived by comparing pencil-beam absorption measurements with a $1\degree$ beam for emission. If the observed column density is significantly below the average column density in the $1\degree$ solid angle, it will produce overestimates of the density and pressure. 

\subsection{Metallicity Effects}
The total ionized gas column density we estimate depends on the gas-phase metallicity, $Z$, as shown in Equation~\ref{eq:SiTot}. As a result, our estimated ionized gas density and thermal pressure are linearly proportional to the metallicity, while the path length is $L~\alpha~\nicefrac{1}{Z^2}$. \citet{Bordoloi2017} estimate the outflowing gas has a subsolar metallicity of of $Z \gtrsim 30\%$ based on photoionization modeling and \ion{O}{1}/\hi\ measurements toward 1H1613-097, a different quasar line of sight passing through the northern Fermi Bubble. Additionally, \citet{Keeney2006} estimated metallicities of $\gtrsim 10-20\%$ solar for other high-velocity clouds toward Galactic Center.

Our measure of the [\nii]/\ha\ line ratio can serve as an independent check of the subsolar metallicity estimate. $\textrm{N}^{+}$ and $\textrm{H}^+$ have similar first ionization potentials of $\unit[14.5]{eV}$ and $\unit[13.6]{eV}$, respectively. As a result, they often exhibit similar ionization levels in photoionized gas such that $\nicefrac{N^+}{N^0} \approx \nicefrac{H^+}{H^0}$ \citep{Haffner1999}. Then the ratio can be expressed 
\begin{equation}
    \textrm{[\nii]/\ha} = \left(1.63~\times10^5\right) \left(\frac{N}{H}\right)~T_4^{0.426}~e^{\nicefrac{-2.18}{T_4}} \label{eq:niiha}
\end{equation}
where $\left(\frac{N}{H}\right)$ is the nitrogen abundance. Photoionization models predict $\nicefrac{N^+}{N^0} \sim 0.8~\times~\nicefrac{H^+}{H^0}$ \citep{Sembach2000}, which would decrease the [\nii]/\ha\ line ratio by $20\%$. 

With a solar nitrogen abundance of $\left( \nicefrac{\textrm{N}}{\textrm{H}} \right)_{\odot} = \left( 6.76 \pm 0.78 \right)~\times~10^{-5}$ \citep{Asplund2009}, and our estimated gas temperature, $T_e = \unit[8500^{+2700}_{-2600}]{K}$, the predicted line ratio from Equation~\ref{eq:niiha} is $\textrm{[\nii]/\ha}= 0.79^{+0.76}_{-0.73}$, with the large error resulting from the large error on $T_e$. If instead, we consider $30\%$ solar nitrogen abundance, then $\textrm{[\nii]/\ha} = 0.24^{+0.23}_{-0.22}$. While both can be consistent with our measured line ratio of $\textrm{[\nii]/\ha} = 0.26 \pm 0.05$ due to the uncertainty in estimated temperatures, our relatively low measured line ratio could better support a subsolar metallicity as suggested in \citet{Bordoloi2017}.

At $30\%$ metallicity, our measured ionized column is $N_{\textrm{H}^+} = \unit[\left(3.28 \pm 0.33\right)~\times 10^{18}]{\textrm{cm}^{-2}}$, resulting in 
\begin{eqnarray*}
  n_{e} & ~=~ & \unit[1.8 \pm 0.6]{\textrm{cm}^{-3}} \\
  L & ~=~ & \unit[0.56 \pm 0.21]{pc} \\
  \nicefrac{p}{k} & ~=~ & \unit[32,000^{+15,000}_{-14,000}]{\textrm{cm}^{-3}~K}.
\end{eqnarray*}

\subsection{Beam Dilution Effects}
In the derivations above, a beam dilution factor of $\epsilon_\textrm{b} = 1$ is used under the assumption that the emitting gas fills the WHAM beam. At the assumed distance of $D \sim \unit[6.5]{kpc}$, the $1\degree$ WHAM beam subtends $\sim \unit[115]{pc}$, but our measured characteristic lengths, $L$, are at least 2 orders of magnitude smaller. Our filled beam assumption is only valid if the emitting gas lies in a very thin sheet, spanning hundreds of parsecs across the sky but only a fraction of a parsec along our line of sight. While this geometry is possible with the \ha\ emission originating in a compressed zone along the Fermi Bubble shell, it is not possible to rule out the possibility of a smaller, higher-density pocket of gas instead of a thin sheet with a single line of sight. As an extreme example, if the emitting gas originated from a sphere with diameter $d = L = \unit[0.56\pm0.21]{pc}$ as estimated with a $30\%$ metallicity, then $\epsilon = \left(2.4\pm1.8\right) \times 10^{-5}$. This results in a true emission measure and electron density that is $42,000 \pm 32,000$ times greater than our estimate. However, we expect that a geometry closer to a thin sheet is most likely based on existing MHD or geometric models \citep[e.g.][see also Figure~\ref{fig:MapArea} and its discussion]{Sarkar2015,Miller2016}.

\subsection{High Positive Velocity Components}\label{sec:meth:upper}
We detect no significant \ha\ or [\nii] emission around $\vlsr \sim \unit[+135]{\kms}$ where UV absorption is observed. 
$3\sigma$ upper limits for \ha\ and [\nii] for this component using an RMS noise of $\unit[0.0015]{R/\left(\kms\right)}$ are both $\unit[0.124]{R}$. 
Since the dust is primarily limited to the foreground in $3$D models, we assume the same $A_V = \unit[1.5 \pm 0.2]{mag}$ from \citet{Green2019}. 
Then assuming the same gas temperature as above, our extinction-corrected emission measure upper limit is $EM < \unit[0.9^{+0.3}_{-0.3}]{\textrm{cm}^{-6}~\textrm{pc}}$. 
Combining with UV absorption column densities from \citet{Bordoloi2017} yields 
\begin{eqnarray*}
  n_e & ~<~ & \left(\unit[4.2 \pm 1.4]{\textrm{cm}^{-3}}\right)~({Z}/{Z_\odot}) \\
  L & ~>~ & \left(\unit[0.11 \pm 0.04]{pc}\right)~({Z}/{Z_\odot})^{-2} \\
  {p}/{k} & ~<~ & \left(\unit[75,000 \pm 34,000]{\textrm{cm}^{-3}~K}\right)~({Z}/{Z_\odot}),
\end{eqnarray*}
where $Z$ is the metallicity.

Bright OH line contamination from the upper atmosphere in our \ha\ spectrum coincides with the UV absorption feature near $\vlsr \sim \unit[+263]{\kms}$. Larger residuals from this line are seen in Figure~\ref{fig:OpticalSpectra} near $\vlsr \sim \unit[+260]{\kms}$. Also, our current [\nii] observations do not extend to this high positive velocity region. We can estimate an approximate $3\sigma$ upper limit for \ha\ emission at the high positive velocity component using an enhanced RMS noise of $\unit[0.015]{R/\left(\kms\right)}$ to be $I_{\ha} < \unit[1.24]{R}$. If the ionized gas density decreases as a function of height above the disk midplane, then the intrinsic emission measure would likely be lower than what is seen for the high negative velocity component. This further reduces our chances to measure any high positive velocity emission from the far side of the Fermi Bubble.

\section{Discussion} \label{sec:disc}

High-velocity ultraviolet absorption-lines detected in sight lines toward the inner Galaxy have previously been interpreted in the context of an outflow associated with the Fermi Bubble. The addition of
optical emission line observations provides a test of this hypothesis.
If the UV absorbing and \ha\ emitting gas originates at a hydrodynamic interface between the outflowing Fermi Bubbles and ambient halo gas, we would expect two key signatures: (1) gas at high pressures, and (2) a bipolar geometry with velocity gradients as a function of latitude and longitude. Our \ha\ observations toward PDS~456 indicate that the first signature is indeed present; the measured thermal pressure is anomalously high for warm ionized gas $\sim \unit[1.3]{kpc}$ above the Galactic midplane. In the solar neighborhood, assuming a midplane density of $n_{e,0}=0.03-0.08~{\textrm{cm}^{-3}}$ and scale height of $1.0-1.8$ kpc, warm ionized gas with temperature $T_e=8000$ K located $\unit[1.3]{kpc}$ above the plane would only have a thermal pressure of $\nicefrac{p}{k}\sim\unit[100-600]{\textrm{cm}^{-3}~K}$. 
\citet{Savage2017} measured a thermal pressure of $\nicefrac{p}{k}\sim\unit[10^{5}]{\textrm{cm}^{-3}~K}$ in an ultraviolet absorption component detected at $\unit[-114]{\kms}$ in the spectrum of LS 4825, a B1 Ib-II star at distance of $\sim\unit[21]{\textrm{kpc}}$. Assuming this absorption arises in the vicinity of Galactic Center, the gas is $\sim \unit[1]{\textrm{kpc}}$ below the Galactic plane.

Our derived pressures of the warm ionized gas are comparable to the thermal pressure predicted by models of hot, X-ray-emitting gas surrounding the Galaxy. A model of the Milky Way's hot halo based on an analysis of \ion{O}{7} and \ion{O}{8} X-ray emission lines \citep{Miller2015} predicts a density and pressure of $n_e = \unit[3 ~\times 10^{-3}]{\textrm{cm}^{-3}}$ and  $\nicefrac{p}{k} = \unit[12,000]{\textrm{cm}^{-3}~K}$ at the modeled distance. It also predicts a pressure of $\nicefrac{p}{k}\sim 54,000{\textrm{cm}^{-3}~K}$ for the high-pressure component seen toward LS 4825, assuming a solar metallicity. In a follow-up work, they incorporated a Fermi Bubble shell model and intended to explain an excess of \ion{O}{7} and \ion{O}{8} X-ray emission in the inner Galaxy \citep{Miller2016}. This work yielded a shell temperature of $\log{\nicefrac{T_e}{K}} = 6.7^{+0.25}_{-0.1}$ and density $n_e = \unit[\left(10 \pm 0.3\right) ~\times 10^{-4}]{\textrm{cm}^{-3}}$, corresponding to a thermal pressure $\nicefrac{p}{k} = \unit[10,000^{+19,000}_{-3000}]{\textrm{cm}^{-3}~K}$ at solar metallicity \citep{Miller2016}\footnote{Since \citet{Miller2016} report hydrogen particle density, we multiply by two to get the total particle density of fully ionized gas.}.
The fraction of their $L_{\textrm{Hot}} = \unit[1]{\textrm{kpc}}$ shell occupied by warm ionized gas with our characteristic length of $L_{\textrm{Warm}} = \unit[0.56 \pm 0.21]{\textrm{pc}}$ is  $\nicefrac{L_{\textrm{Warm}}}{L_{\textrm{Hot}}} = \left(5.6 \pm 2.1\right)\times10^{-4}$. 

The concordance between our thermal pressure estimates and the expected thermal pressure of a hot gas halo supports the hypothesis that the observed absorption is associated with warm ionized gas above the central Galaxy. Uncertainties in our thermal pressure measurements, combined with the uncertainties in the pressure estimate of the hot halo and Fermi Bubble shell, do not allow us to say whether the gas arises in the bipolar shell of a nuclear outflow or from gas embedded in a hot medium, similar to the observed \hi\ \citep{McClure-Griffiths2013, DiTeodoro2018, Lockman2020}. 

\begin{figure*}[]
\epsscale{1}
\plotone{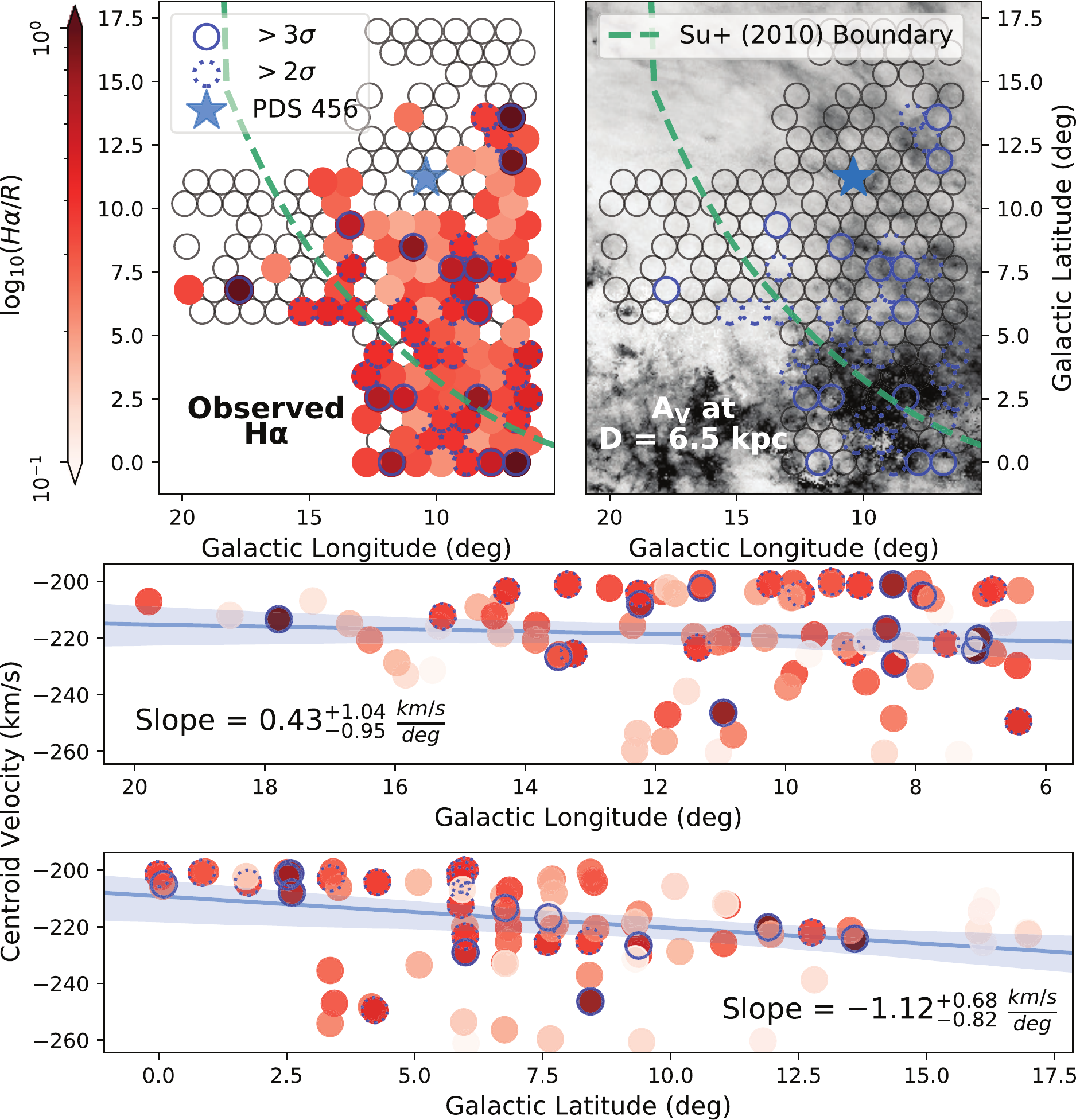}
\caption{Preliminary \ha\ map surrounding the PDS~456 quasar composed of 120s exposures per pointing integrated between $\unit[-270]{\kms} < \vlsr < \unit[-200]{\kms}$) (\emph{top left}). Only pointings above a $1\sigma$ detection threshold are colored by their intensity, while open circles show the location of pointings with no detected emission. $2\sigma$ and $3\sigma$ detections are further outlined with blue, dotted and solid outlines, respectively. Extinction estimates from \citet{Green2019} are shown out to a distance of $\unit[6.5]{kpc}$, with darker regions indicating more dust (\emph{top right}). A green dashed line marks the approximate outline of the northern Fermi Bubble from \citet{Su2010}. Velocity centroids as a function of Galactic longitude (\emph{middle}) and Galactic latitude (\emph{bottom}) are shown for measurements with errors $< \unit[12]{\kms}$. On average, the plotted velocity centroids have standard errors of $\unit[8]{\kms}$. Robust least-squares linear-regression fits are shown in blue with $95\%$ confidence intervals from $1000$ bootstrap resamples. \label{fig:MapArea}}
\end{figure*}

While UV absorption-line observations are limited by the availability of background targets and 21-cm observations only probe sparsely distributed neutral gas, mapping extended optical line emission has the potential to trace the continuity of gas kinematics above and below the inner Galaxy.  
To this end, we initiated a campaign to spectroscopically map the sky at high velocities surrounding the footprint of the Fermi Bubbles. In Figure~\ref{fig:MapArea}, we show early results from mapping a small region surrounding PDS~456 at high negative velocities. We see evidence for extended emission at high negative velocities in the vicinity of PDS~456, shown in the top left panel. The top right panel shows the estimated extinction in $A_V$ from \citet{Green2019} out to a distance of $\unit[6.5]{kpc}$. 

Most \ha\ detections lie in regions with substantial foreground dust. If high-velocity \ha\ were pervasive in the inner Galaxy, the relatively low dust columns for sight lines with $l>15\degree$ and $b > 5\degree$ mean that the \ha\ emission should be brighter and easier to detect.  The lack of observed emission in this region combined with the presence of \ha\ emission in dustier directions closer to the Galactic Center suggests that the high-velocity emission might be due to a coherent central structure, possibly related to the Fermi Bubble. 

Velocity centroids of the emitting gas are shown in the lower panel. These velocity estimates have large errors and currently show no strong evidence for a gradient as predicted in kinematic outflow models \citep[e.g.][]{Bordoloi2017, DiTeodoro2018} or seen in \hi\ observations \citep{Lockman2020}. Ultimately, this initial map is composed of low signal to noise data and cannot be used to draw definitive conclusions. Future work and continued WHAM observations will allow this initial map to be expanded to better confront model predictions and identify boundaries.

In the future, our \ha\ observations close to, or from within, the boundaries of the Fermi Bubble can also provide an opportunity to constrain the radiation field emerging from Galactic Center. Previously, \ha\ observations from WHAM of ionized gas in the Magellanic stream have been used to to diagnose the radiation field emerging from Galactic Center $\sim\unit[4]{\textrm{Myr}}$ ago \citep{JBH2013, Barger2017}. The smaller distance to Galactic Center for the gas we discuss here would provide a much more recent glimpse into the ionizing radiation field emerging from Sgr A* and the CMZ, as well as more localized ionization sources, such as shocks or cooling flows.

\section{Conclusions} \label{sec:conc}
We have discovered high-velocity optical emission in the vicinity of the $2$D footprint of the Fermi Bubbles at the same velocity as previously observed UV absorption features 
\citep{Fox2015}. As a result, we have measured model-independent constraints on the in situ physical conditions of warm gas above Galactic Center toward the quasar PDS~456. We summarize our findings to be: 
\begin{enumerate}[itemsep=0mm,leftmargin=*]
    \item \ha\ and [\nii] emission are detected at high negative velocity with extinction-corrected intensities of $I_{\ha} = \unit[0.84^{+0.10}_{-0.09}]{R}$ and $I_{\left[\textrm{N\rom{2}}\right]} = \unit[0.225^{+0.028}_{0.025}]{R}$.
    \item Optical emission and UV absorption from low ions have line widths indicating a gas temperature of $T_e = \unit[8500^{+2700}_{-2600}]{K}$ with a nonthermal contribution of $\sigma_{nonT} = \unit[10.5^{+1.5}_{-1.6}]{\kms}$.
    \item The observed optical line ratio of $\textrm{[\nii]/\ha} = 0.26 \pm 0.05$ is consistent with abundances of approximately $30\%$ solar when compared to an estimate of the gas metallicity toward a different sight line probing the northern Fermi Bubble \citep{Bordoloi2017}. 
    \item Assuming $30\%$ solar metallicity, the ionized gas has a characteristic density and length of $n_{e} = \unit[1.8 \pm 0.6]{\textrm{cm}^{-3}}$ and $L_{e} = \unit[0.56 \pm 0.21]{pc}$ with a thermal pressure of $\nicefrac{p}{k} = \unit[32,000^{+15,000}_{-14,000}]{\textrm{cm}^{-3}~K}$. This high thermal pressure is comparable to, but still greater than, those predicted by models of a hot gas halo in the inner Galaxy or of a Fermi Bubble shell \citep{Miller2015,Miller2016}. 
    \item Initial \ha\ spectroscopic maps of the region surrounding PDS~456 reveal extended emission at the same high negative velocities. 
\end{enumerate}

With future observations, WHAM can trace emission associated with the Fermi Bubbles both spatially and kinematically at large scales. Additionally, other pointed observations toward distant UV bright sources with existing HST spectra can provide sensitive column-density profiles of multiple species across different regions of the southern and northern Fermi Bubbles.

\acknowledgments
We thank Andrew Fox for providing useful comments and constraints on the UV absorption-line measurements and total ionized gas column density. We acknowledge the support of the U.S. National Science Foundation (NSF) for WHAM development, operations, and science activities. The optical observations and work presented here were funded by NSF awards AST-0607512, AST-1108911, and AST-1714472/1715623/1940634. R.A.B would like to acknowledge support from NASA grant NNX17AJ27G. Some of this work took part under the program SoStar of the PSI2 project funded by the IDEX Paris-Saclay, ANR-11-IDEX-0003-02. The authors acknowledge Paris-Saclay University's Institut Pascal program "The Self-Organized Star Formation Process" and the Interstellar Institute for hosting discussions that nourished the development of the ideas behind this work. This work uses observations made with the NASA/ESA Hubble Space Telescope, obtained from the Data Archive at the Space Telescope Science Institute, which is operated by the Association of Universities for Research in Astronomy, Inc., under NASA contract NAS5-26555. These observations are associated with program 13448.

\ \vspace{1ex}

\facilities{WHAM, HST (COS)}

\software{\texttt{astropy} \citep{astropy},
        \texttt{matplotlib} \citep{mpl}, 
        \texttt{seaborn} \citep{sns}, 
        \texttt{whampy} \citep{whampy},
        \texttt{lmfit} \citep{lmfit}, 
        \texttt{dustmaps} \citep{dustmaps}, 
        \texttt{bettermoments} \citep{Teague2018},
        \texttt{emcee} \citep{emcee}.
          }


 \newcommand{\noop}[1]{}


\end{document}